\documentclass[aps,pra,twocolumn]{revtex4-1}
\usepackage[T1]{fontenc}
\usepackage{lmodern}
\usepackage{amsmath}
\usepackage{amssymb}
\usepackage{graphicx}
\usepackage{multirow}
\usepackage{amsfonts}
\providecommand{\U}[1]{\protect\rule{.1in}{.1in}}
\begin{document}
\title{Theory of Mn-doped I-II-V Semiconductors}
\author{J.~K.~Glasbrenner,$^{1}$ I.~\v{Z}uti\'c,$^{2}$ and I.~I.~Mazin$^{3}$}
\affiliation{$^{1}$ National Research Council/Code 6393, Naval Research Laboratory,
Washington, DC 20375, USA}
\affiliation{$^{2}$ Department of Physics, University at Buffalo, State University of New
York, NY 14260, USA}
\affiliation{$^{3}$Code 6393, Naval Research Laboratory, Washington, DC 20375, USA}
\date{\today}

\begin{abstract}
A recently discovered magnetic semiconductor Ba$_{1-x}$K$_{x}$(Zn$_{1-y}$Mn$_{y}$)$_2$As$_{2}$,
with its decoupled spin and charge doping, provides a unique opportunity to
elucidate the microscopic origin of the magnetic interaction and ordering in
dilute magnetic semiconductors (DMS). We show that (i) the conventional
density functional theory accurately describes this material, and (ii) the
magnetic interaction emerges from the competition of the short-range
superexchange and the longer-range spin-spin interaction mediated by the
itinerant As holes. The latter can be viewed as a high-doping extrapolation of
with the Schrieffer-Wolff $p-d$ interaction representing an
effective Hund's rule coupling, $J_{H}^{\text{eff}}$. The key difference
between the classical double exchange and the actual interaction in DMS is
that an effective $J_{H}^{\text{eff}}$, as opposed to the standard Hund's
coupling $J_{H}$, depends on the Mn $d-$band position with respect to the
Fermi level, and thus allows tuning of the magnetic interactions. The physical
picture revealed for this transparent system may also be applicable to more
complicated DMS systems.
\end{abstract}
\maketitle

\emph{Introduction}-The dilute magnetic semiconductors (DMS) are nonmagnetic
semiconductors doped with magnetic elements and displaying various
manifestations of magnetic
ordering~\cite{Furdyna1988:JAP,Ohno1998:S,Dietl2010:NM,Jungwirth2006:RMP,Zutic2004:RMP}. The
carrier-mediated magnetism in DMS offers a versatile control of the exchange
interaction by tuning the Curie temperature $T_{C}$ through changes in the
carrier density, for example by an applied electric field, photoexcitations, or
even heating~\cite{Zutic2004:RMP,Koshihara1997:PRL,Ohno2000:N,Petukhov2007:PRL}.
However, despite the four decades of intensive work on DMS,  challenges remain
and materials complexity often hinders theoretical understanding. The origin of
magnetic
ordering~\cite{Furdyna1988:JAP,Ohno1998:S,Dietl2010:NM,Zutic2004:RMP} 
and paths to higher $T_{C}$ remain strongly 
debated \cite{Dietl2010:NM,Tanaka2014:APR,Dobrowolska2012:NM}.

The Mn-doping of II-VI and III-V semiconductors is the usual method for
synthesizing DMS. In the II-VI DMS Mn$^{2+}$ is isovalent with the group II
ions and provides only spin doping;  the lack of carriers makes robust
ferromagnetism elusive. In the III-V compounds introducing Mn leads to both
spin and carrier doping, but a low-solubility limit for Mn complicates growth
and can lead to nanoscale clustering of Mn ions. This dual role of  Mn
complicates theoretical understanding and creates difficulties in establishing
the connection between host properties and figures of merit. For example, the
$T_{C}$ for (Ga,Mn)N is predicted to be $T_{C}>300$ K~\cite{Dietl2000:S}, but
in experiment it is much lower, $T_{C}\lesssim10$ K~\cite{Sawicki2010:PRB}.
Finallly, both substitutional and interstitial Mn are thermodynamically stable
and form during synthesis, which additionally complicates theoretical treatment.

The recent discovery of the I-II-V DMS
compounds~\cite{liznas,Deng2013:PRB,Zhao2013:NC} provides a way to overcome
these difficulties. In contrast to the II-VI and III-V compounds, in the I-II-V
ones hole and spin doping are controlled separately by substitution with the
group I and group II ions, respectively. In (Ba$_{0.7}$K$_{0.3}%
$)(Zn$_{0.85}$Mn$_{0.15}$)$_{2}$As$_{2}$, a $T_{C}\sim220$ K~\cite{Zhao2014:CSB}
is already higher than $\sim190$ K~\cite{Dietl2010:NM} attained in (Ga,Mn)As,
the prototypical III-V DMS. Unlike (Ga,Mn)As, both $p$- and $n$-doped I-II-Vs
can be ferromagnetic~\cite{Zhao2013:NC,Man2014:P}, and a coercive field
$\sim10^{4}$ Oe in (Ba,K)(Zn,Mn)$_{2}$As$_{2}$ at 2 K~\cite{Zhao2013:NC} is two
orders of magnitude larger than in (Ga,Mn)As. Apart from potential
applications~\cite{Zutic2004:RMP}, the I-II-V DMS compounds are well suited for
theoretical study, because (1) the Mn$^{2+}$ is isovalent with Zn, (2) charge is
doped into the Ba sublayer, spatially and electronically disconnected from the
active (Zn,Mn)$_{2}$As$_{2}$ layers, and (3) interstitial locations for Mn ions
are energetically precluded.

A key feature that a theory of I-II-V DMS materials must capture is the
curious result that despite the high-temperature measurements indicating a
high spin state with 5 $\mu_{B}$/Mn, the low-temperature ferromagnetic
magnetization measurement finds moments of $\lesssim2\text{ }\mu_{B}$. The
authors of Ref.~\cite{Zhao2013:NC} conjectured that this reduction may be due
to the formation of nearest neighbor Mn$_{2}$ singlets. We will show below
that this is a plausible explanation, noting that a correct theory should
explain why singlets form and estimate their concentration.

In this Letter, we present density functional theory (DFT) calculations of the
energetics of Mn pairs substituted in the Zn lattice as a function of the pair
distance and hole-doping with K. We extract exchange parameters from these
calculations and find that ordering changes from antiferomagnetic (AFM) with no
hole doping to ferromagnetic (FM) with hole doping, with the exception of
nearest neighbor (nn) pairs, which remain AFM and are energetically preferred.
We then show using thermodynamic arguments that singlet formation is responsible
for the reduced magnetization in Ref.~\cite{Zhao2013:NC}. We also address the
different terminologies used for the effective magnetic interaction between the
Mn $d$ and As $p$ states, such as double exchange~\cite{Khomskii:2010}, the
Zener $p-d$ model~\cite{Zener1951b:PR}, and the RKKY
interaction~\cite{rkky1,rkky2,rkky3}. In our view, these all describe the same
indirect exchange interaction \cite{Nagaev:1983} and therefore the same basic
physics.

\begin{figure}[ptb]
\begin{center}
\includegraphics[width=0.48\textwidth]{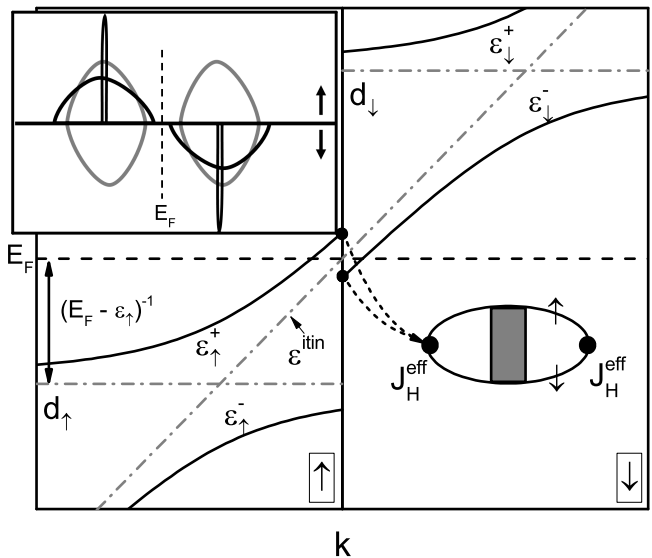}
\end{center}
\caption{A cartoon of the interaction between localized spins
mediated by the itinerant carriers depicting a spin-resolved schematic of the broad
As band $\varepsilon^{\text{itin}}$ hybridizing with the narrow Mn bands
$d_{\uparrow}$ and $d_{\downarrow}$, forming the bands $\varepsilon^{\pm
}_{\uparrow,\downarrow}$. The effective magnetic coupling J$_{H}^{\text{eff}}$
scales with $\left(  E_{F} - \varepsilon_{\uparrow} \right) ^{-1}$. Inset: A
schematic of the BaZn$_{2}$As$_{2}$ DOS before (gray lines) and after (black
lines) Mn doping. The narrow Mn states interact with the As states, broadening
the DOS and reducing the indirect band gap.}
\label{pic-bandsanddos}
\end{figure}

\textit{Calculations}-We employed two DFT implementations: a pseudopotential
method (VASP \cite{Kresse1993:PRB,Kresse1996:PRB}) and a full potential linear
augmented plane wave (LAPW) method (ELK \cite{elk}). Selected results have
been verified against an alternative LAPW package, WIEN2k \cite{wien2k}. A
generalized gradient approximation \cite{Perdew1996:PRL} was used for total
energy calculations and the modified Becke-Johnson
functional~\cite{Becke2006:JCP,tranblaha} (known to give correct band gaps for
semiconductors) was used for analyzing the electronic structure. For pure BaZn$_{2}%
$As$_{2}$ we obtain the indirect gap of $0.25\text{ eV}$ between the $\Gamma$
and $Z$ points, and the direct gap of $0.71\text{ eV}$ at the $\Gamma$ point,
in agreement with previous calculations \cite{bza-dft}, see Supplementary Material.

We first analyze the effect of Mn doping (Ba(Zn$_{1-y}$Mn$_{y}$)$_2$As$_{2},$
$y>0$). The inset in Fig.~\ref{pic-bandsanddos} qualitatively illustrates the
spin-resolved density of states (DOS) (see Supplementary Material for explicit
DOS calculations). There are five Mn bands in each spin, confirming the
Mn$^{2+}$ state. The calculated Mn moment is $\sim4.7\ \mu_{B}$ in ELK and
$\sim4.9\ \mu_{B}$ in VASP~\cite{partition}. The valence band is predominantly
As $p$ states, the conductance band Ba states. The As states hybridize with Mn
as depicted in Fig.~\ref{pic-bandsanddos}, which is parameterized by hopping
parameter $t_{pd}$ \cite{hop}. As a result, the top of the As spin-majority band
is pushed up and the bottom down at a rate of $\sim2.8y$ eV, so for $y=0.25$,
there is a shift of $0.7$ eV. In contrast, the bottom of the conductance (Ba)
spin-minority band is pushed down because of the hybridization with unoccupied
Mn states. This provides the magnetic coupling between the local spin and
itinerant carriers \cite{Cibert:2008}. Another manifestation of the same effect,
verifiable experimentally, is that with Mn doping the indirect gap between the
top of the spin-up valence band and the bottom of the spin-down conduction band
is reduced and eventually closes when the doping is large enough, see the inset
of Fig.~\ref{pic-bandsanddos}.

By introducing hole doping [Ba$_{1-x}$K$_{x}$(Zn$_{1-y}$Mn$_{y}$)$_2$As$_{2},$
$x>0$], the calculated Mn moments change. At $x=0.4$ they are reduced by 40\%
and As atoms acquire opposite moments (Ba and Zn remain unpolarized). Both
effects are caused by the Mn-As hybridization, while K doping makes the effect
visible. Indeed, because of the upshift of the top of the As band, there are
more holes in the spin-majority band, creating negative polarization on As.
Furthermore, because of proximity the hybridization of As holes with the
spin-majority Mn states is stronger than with the spin-minority ones, so holes
carry more spin-majority Mn character and hole-doping reduces the Mn moments.

Next we constructed different supercells, placing Mn pairs into different
substitutional positions. Unlike (Ga,Mn)As, where Mn easily occupies
interstitials, complicating the theoretical analysis, in BaZn$_{2}$As$_{2}$ this
is essentially impossible. The calculated free energy penalty for interstitial
vs substitutional Mn doping is huge, $F_{int}-F_{sub}>2.4$ eV/Mn, for all
admissible values of the Zn chemical potential (see Supplementary Material).

\begin{figure}[ptb]
\begin{center}
\includegraphics[width=0.48\textwidth]{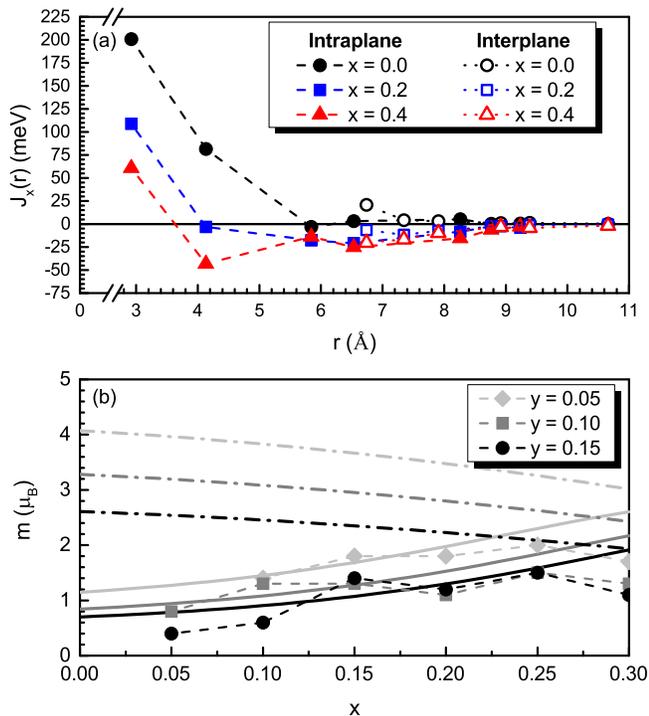}
\end{center}
\caption{(Color online) The magnetic interaction's doping dependence. (a) The
exchange coupling $J_{x}(r)$ as a function of distance between Mn pairs for
different hole dopings $x$. (b) The dependence of the reduced magnetization
due to singlet formation on hole doping. The symbols are the experimental
measurements from Ref.~\cite{Zhao2013:NC}. The solid lines are the reduced
magnetization when Mn singlet formation is in thermodynamic equilibrium at
$500$ K, while the dash-dot lines are the reduced magnetization for a random
distribution of Mn atoms. The light gray lines correspond to $y = 0.05$, the
dark gray lines to $y = 0.10$, and the black lines to $y = 0.15$.}%
\label{pic-exchangeandmag}%
\end{figure}

We now assume a Heisenberg model for the Mn-Mn magnetic interactions at
lattice sites $i,j$,
\begin{equation}
	\label{eq-Heiss} H=\sum_{i<j}J_{x}^{ij}{\hat{S}}_{i}\cdot{\hat{S}}_{j},
\end{equation}
where the $\hat{S}_{i,j}$ are the unit vectors in the spin directions. We can
map~\cite{map} the calculated energies for different magnetic configurations
onto Eq.~(\ref{eq-Heiss}) and extract the spatial dependence of the exchange
$J_{x}^{ij}$. 

Figure \ref{pic-exchangeandmag}(a) summarizes $J_{x}^{ij}$ for both intraplanar
and interplanar Mn pairs up to 7 neighbors for $x=0$, $0.2$, and $0.4$ hole
dopings (for more details, see the Supplementary Material). We note that the
intraplanar and interplanar results roughly lie on the same universal curve, so
we define $J_{x}(r) \equiv J_{x}^{ij}$. Without hole doping ($x=0$), $J_{x}(r)$
is AFM for all pairs and decays strongly with distance, consistent with
superexchange~\cite{Nagaev:1983, Khomskii:2010} .

Hole doping drives the system toward ferromagnetism, so that $J_{x}(r)$ becomes
FM for 2nd and higher neighbors. For nn pairs, $J_{x}(r)$ remains AFM even for
$x=0.4$, but is reduced threefold. This reduction, along with the 2nd
neighbor's exchange parameter barely changing sign to become FM for $x=0.2$ ,
reveals that this behavior is due to the competition between the short-range
AFM superexchange and a longer-range FM interaction.

We now address the puzzling reduction of the net magnetization $M$ compared with
the local Mn magnetic moments. We verified that even for $x=0.4$ doping that the
nn exchange parameter is AFM, such that nn Mn pairs form a singlet. Let us first
assume that Mn dopants are randomly distributed in the Zn lattice and estimate
the magnetization reduction. If we neglect clusters of 3 or more Mn atoms (i.e.,
assuming $y\ll1$),  we obtain $M_{\text{theor}}(x,y)=M(x)(1-y)^{4}$. Depending
on $x,$ $M(x)\approx3$--$5\text{ }\mu_{B},$ and we can interpolate $M(x)$ using
our results. Figure \ref{pic-exchangeandmag}(b) shows $M_{\text{theor}}(x,y)$
for three values of $y$ (dash-dotted lines) and compares it with the experiment
of Ref.~\cite{Zhao2013:NC}. While there is some qualitative agreement, the
magnetization suppression is noticeably underestimated, especially for small
$x.$

Our calculations also indicate an energetic preference for nn Mn pairs to form.
This, along with the variation in experimental Mn moments, leads us to conclude
that the suppression is sensitive to sample preparation and hence the Mn
distribution is not entirely random. To quantify this, we use the calculated
energy differences between the AFM nn pair and a remote FM pair: $\Delta
E(x)\sim185,$ 80 and $-30$ meV, for $x=0,$ 0.2 and 0.4, respectively. We can now
evaluate the free energy using these values and the combinatorial entropy (see
the Supplementary Material) to obtain the moment reduction
$r=(\sqrt{\beta^{2}+16\beta y}-\beta)/8y$, correct in the $y\ll1$ limit, where
$\beta=\exp[-\Delta E/T]$. Interpolating the calculated $\Delta E(x)$ and
$M(x)$, and using as the effective synthesis temperature 500 K~\cite{temp}, we
get the solid lines shown in Fig.~\ref{pic-exchangeandmag}(b). Given the
variation in the experimental data and the lack of any adjustable parameter, the
agreement is excellent. From these calculations we can predict that quenching,
rather than slow cooling, may be advantageous for enhancing the magnetization
per Mn, which can be as high as 3 $\mu_{B}.$

\emph{Discussion}-We determined that the magnetic ordering is a combination of
a short-range AFM interaction and a longer-range FM interaction. We identify
the AFM interaction as superexchange, which is compatible with Mn being in the
high-spin $S=5/2$ state, and is accounted for in DFT. The basic picture of
superexchange is that there is an effective amplitude $\tilde{t}_{dd}$ for a
$d-$electron to hop from one Mn to another. For nn pairs $t$ is large, as only one
intermediate hop to an As $p$ state is required, and in addition there is some
direct overlap between Mn $d_{x^{2}-y^{2}}$ orbitals. If the alignment is FM,
hopping leads to a splitting of the occupied Mn states with no gain in kinetic
energy. For an AFM alignment, where hopping proceeds from occupied to
unoccupied states, this leads to a downshift of the occupied Mn states by
2$\tilde{t}_{dd}^{2}/U,$ where $U$ is the energy cost of flipping the spin of
one $d$ electron. This cost in DFT is $\sim5J_{H},$ where $J_{H}\sim0.8$ eV is
the Hund's rule coupling in Mn (in the Hubbard model $U$ comes from the Coulomb
repulsion and may be larger than $5J_{H})$. This creates the superexchange
coupling $J_{\text{SE}}\approx2\tilde{t}_{dd}^{2}/U$. For farther neighbors
the hopping probability involves multiple hoppings via high-lying Zn states
and rapidly decays.

We will argue now~\cite{akaiargument} that the long-range FM ordering is a
version of double exchange (DE)~\cite{Khomskii:2010}, but first we give an
overview of the DE interaction. The original model~\cite{Zener1951b:PR} assumed
a strong Hund's coupling between localized spins and itinerant electrons from
the same atomic species, which in practice is due to non-integer valency, and
has led to the misconception that DE itself \emph{requires} mixed valency.
Instead, the only real requirement is that the interaction of the local spins
with itinerant electrons be described by an operator of the form
$J_{H}\hat{S}\cdot\hat{\sigma}$. Note that the nature and sign of $J_{H}$ does
not matter, because in the end $J_{H}$ is squared.

The other essential ingredient of DE was the itinerant carriers delocalizing to
lower their kinetic energy, which preferred a FM arrangement of the local spins.
In the original model~\cite{Zener1951b:PR} the strong coupling limit
$J_{H}\rightarrow \infty$ was assumed in order to simplify the calculations, but
that was not a necessary condition for DE. The DE picture, that of itinerant
electrons adjusting their spin density to the background of local spins with
some configurations being more energetically preferable, is simply the standard
spin response theory described in the weak coupling regime by the linear spin
susceptibility $\chi(q)$. In general $\chi(q)$ depends on the electronic
structure and Fermi surface geometry, so again for simplicity it is often
approximated by its value at the $\Gamma$ point, $\chi(q)=\chi(0)=N_{\uparrow
}(0)$, which is not a bad approximation when all sites contain a local moment.
Again, we note that this approximation is not essential when defining DE.
Finally, if the concentration of itinerant carriers is small such that the Fermi
length $2\pi/k_{F}\gg d$ ($d$ is the average distance between spins), then the
response is FM for all relevant distances. For larger $d$ the response decays
rapidly and might acquire an oscillatory part, which depends on the Fermi
surface. This is known as the RKKY interaction \cite{rkky1,rkky2,rkky3}.

To review, the general picture is that the DE implies a local spin interacting
via exchange with an itinerant sea of carriers, which itself responds by
adjusting its spin density to align with the other localized spins, and then
another of these localized spins interacts via exchange with the sea. There are
two elementary exchange processes involved, see the diagram in
Fig.~\ref{pic-bandsanddos}, hence DE. In other words, DE and RKKY are two
different sides of the same coin.

With this clarified, we now turn to the details of the effective exchange
interaction between the local Mn spins and itinerant As holes. As discussed, in
DFT the As electrons at the Fermi level hybridize with Mn and acquire
spin-splitting, see Fig.~\ref{pic-bandsanddos}. The upshift of spin-majority
states at the Fermi energy is $5Zt_{pd}^{2}/(E_{F}%
-\varepsilon_{\uparrow})$ (the bottom of the As band shifts \emph{down}), where
$Z=4$ is the Mn-As coordination number. Here the spin-minority Mn states are
located at $\varepsilon_{\downarrow}>E_{F},$ and the spin majority ones at
$\varepsilon_{\uparrow}<E_{F}$, and, for simplicity, the hopping amplitude
$t_{pd}$ is assumed to be the same for all Mn $d$ states. Similarly, the spin
minority bands are shifted down by $5Zt_{pd}^{2}/(\varepsilon_{\downarrow
}-E_{F})$. This yields an effective Mn spin splitting and thus an effective
Hund's rule coupling of
\[
J_{H}^{\mathrm{eff}}=\frac{-Zt_{pd}^{2}(\varepsilon_{\uparrow}-\varepsilon
_{\downarrow})}{(E_{F}-\varepsilon_{\uparrow})(\varepsilon_{\downarrow}%
-E_{F})}=\frac{-Zt_{pd}^{2}U}{(E_{F}-\varepsilon_{\uparrow})(\varepsilon
_{\uparrow}+U-E_{F})}.
\]
This is formally the same as the Schrieffer-Wolff transformation frequently used
for Kondo systems \cite{Cibert:2008,Khomskii:2010}. The DMS literature typically
refers to this as the $p-d$ model. We emphasize that the $p-d$
model~\cite{Nagaev:1983} is not an alternative to DE, but a modification of the
latter, where $J_{H}$ is replaced with $J_{H}^{\text{{eff}}}$. The RKKY theory
is in the same spirit, modifying the same physics in a different way by lifting
the $q=0$ approximation, in weak coupling, and using a $q$-dependent
susceptibility. In all cases the effective coupling appears as a pair of
vertices attached to a polarization bubble as in Fig.~\ref{pic-bandsanddos}, and
so the sign of $J_{H}^{\text{{eff}}}$ is irrelevant.

Unlike $J_{H,}$ in principle $J_{H}^{\text{{eff}}}$ is tunable by changing the
spin-flip energy cost $U$ and the position of the occupied $d$-level
$\varepsilon_{\uparrow}$. In our DFT calculations $E_{F}-\varepsilon
_{\uparrow}\sim U/2$, which is the least advantageous situation. We suggest that
substituting As with Sb or P may shift $\varepsilon_{\uparrow}$ up or down,
yielding a $E_{F}-\varepsilon_{\uparrow}$ closer to $U/5$ or $4U/5$ and
increasing $J_{H}^{\text{{eff}}}$ by $\sim60\%$. Assuming that other parameters
remain unchanged, an enhancement of exchange coupling could increase $T_{C}$ by
a factor of $2.5$, which suggests a path forward to room-temperature FM ordering
in the I-II-V compounds.

\emph{Conclusions}-We have shown, based on our first principles calculations,
that ferromagnetism in (Ba,K)(Zn,Mn)$_{2}$As$_{2}$ is a result of the
interaction of localized Mn spins with itinerant As holes that have a
ferromagnetic spin response for all relevant Mn-Mn distances, except for nearest
neighbors. This is a variant of the classical double exchange with the simple
modification of replacing the Hund's coupling $J_{H}$ by the effective $p-d$
coupling $J_{H}^{\text{eff}}$.

The nearest neighbor magnetic interaction is a combination of Anderson's
superexchange that is weakened, but not overcome, by the ferromagnetic double
exchange, and for a K concentration less than $\sim0.35$ it is energetically
advantageous for Mn to form nearest neighbor singlet pairs. Our calculations
describe this process quantitatively and predict a net magnetization reduction
from the ideal 5 $\mu_{B}$/Mn in excellent agreement with experiment.

While our findings have focused on the (Ba,K)(Zn,Mn)$_{2}$As$_{2}$ compound, we
believe that the transparent and simple physical picture that has emerged from
studying this unique system is more general and applicable to other DMS
compounds. Our theory and calculations are uncomplicated by multiple chemical
issues common to other generations of DMS compounds such as Mn-doped IV-VI,
III-V, and II-VI materials. Thus, this new generation of I-II-V materials is an
exciting playground for experimentalists and theorists alike and deserves
further study to elucidate the intrinsic physics of DMS materials.

\emph{Acknowledgments}-We are grateful to A.~Petukhov, F.~Ning, and C.~Q.~Jin
for their useful discussions. I.I.M.~acknowledges funding from the Office of
Naval Research (ONR) through the Naval Research Laboratory's Basic Research
Program. J.K.G.~acknowledges the support of the NRC program at NRL. I.~Z.~was
supported by DOE-BES Grant No.~DE-SC0004890 and ONR.

\appendix

\section*{Supplementary Material}
\label{sect-appendix}

\begin{figure*}
\begin{center}
\includegraphics[width=0.80\textwidth]{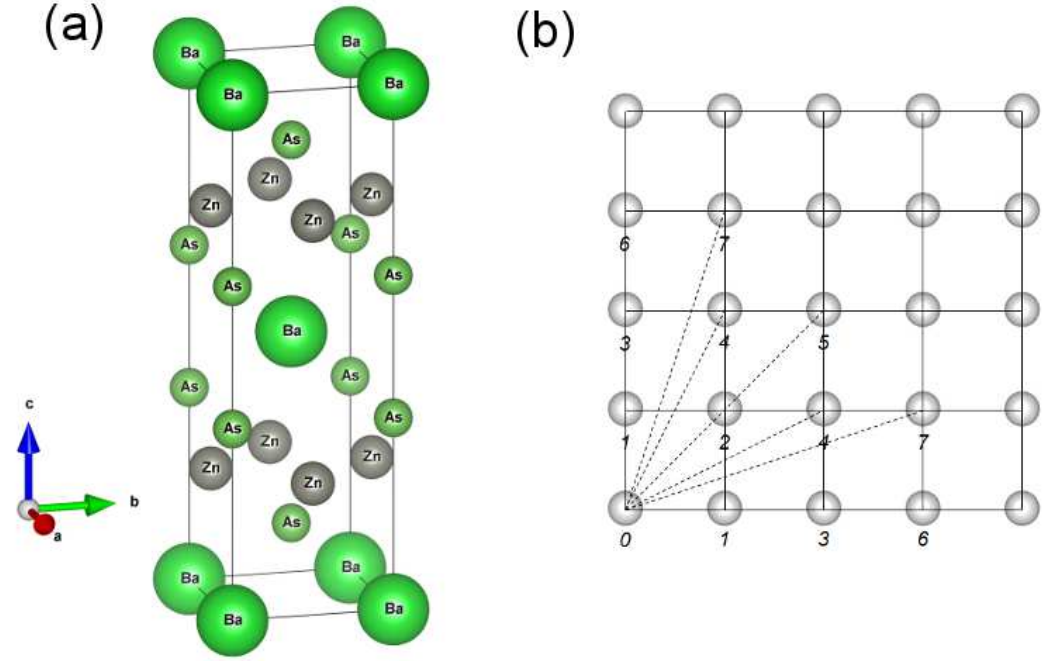}
\end{center}
\caption{(Color online) (a) The crystal structure of the tetragonal
ThCr$_2$Si$_2$ phase of BaZn$_2$As$_2$. (b) A
schematic view of the two-dimensional Zn planes. The labels indicate the
neighbor numbers with one Mn fixed at
site 0 and another placed at sites 1-7. The nearest neighbor interaction
connects sites 0 and 1, the 2nd nearest neighbor interaction connects
sites 0 and 2, and so on. For the interplane interactions, the nearest
interplane neighbor interaction connects site 0 in one plane with
site 0 in the neighboring plane, the 2nd nearest interplane neighbor interaction
connects site 0 in one plane with site 1 in the neighboring plane,
and so on.} \label{pic-bza}
\end{figure*}

\emph{Additional computational details}-A schematic of the crystal structure of
BaZn$_2$As$_2$ is depicted in Fig.~\ref{pic-bza}(a). BaZn$_2$As$_2$ belongs to
space group $I4/mmm$, and for our calculations we used the experimental lattice
constants $a = 4.131 \text{ \AA}$ and $c = 13.481 \text{ \AA}$. These constants
are taken from the experiment in Ref.~\cite{Zhao2013:NC} and correspond to
the co-doped (Ba,K)(Zn,Mn)$_2$As$_2$ system. The Wyckoff sites for the atoms are
$2a$, $4d$, and $4e$ for Ba, Zn, and As respectively, and the internal parameter
for As is set to $z_{\text{As}} = 0.3645$. For all supercells of BaZn$_2$As$_2$
with Mn pairs substituted for Zn atoms, we relax the internal atomic positions
in VASP for the ferromagnetic (FM) and antiferromagnetic (AFM) magnetic
configurations. The $a$ and $c$ lattice parameters are not relaxed, as the
doping in our system is light and in an experimental system would have a
negligible effect on these parameters.

The virtual crystal approximation (VCA) was used to simulate hole-doping (in
experiment it is the substitution of K for Ba), and the implementation depends
on the code used. The VCA in ELK is implemented in the standard way, by
introducing a fictitious atom at the Ba sites which has a fractional charge
between that of Cs and Ba, such as $Z = 55.6$. For VASP, the VCA corresponds to
a weighted average of the pseudopotentials for Ba and K, such as 80\% Ba and
20\% K. The electronic structure generated using this method is consistent with
the electronic structure obtained using the VCA in ELK. We note that atomic
relaxation is not possible when using the VCA in VASP, so we take the relaxed
structure from the Ba(Zn,Mn)$_2$As$_2$ supercells. The use of the VCA in both
ELK and VASP is reasonable as there are no Ba states near the Fermi energy and
this approach has been used successfully on the similar compound BaMn$_2$As$_2$
\cite{glasbrenner} as well as in isostructural Fe-based superconductors.

\begin{figure*}
\begin{center}
\includegraphics[angle=270,width=0.80\textwidth]{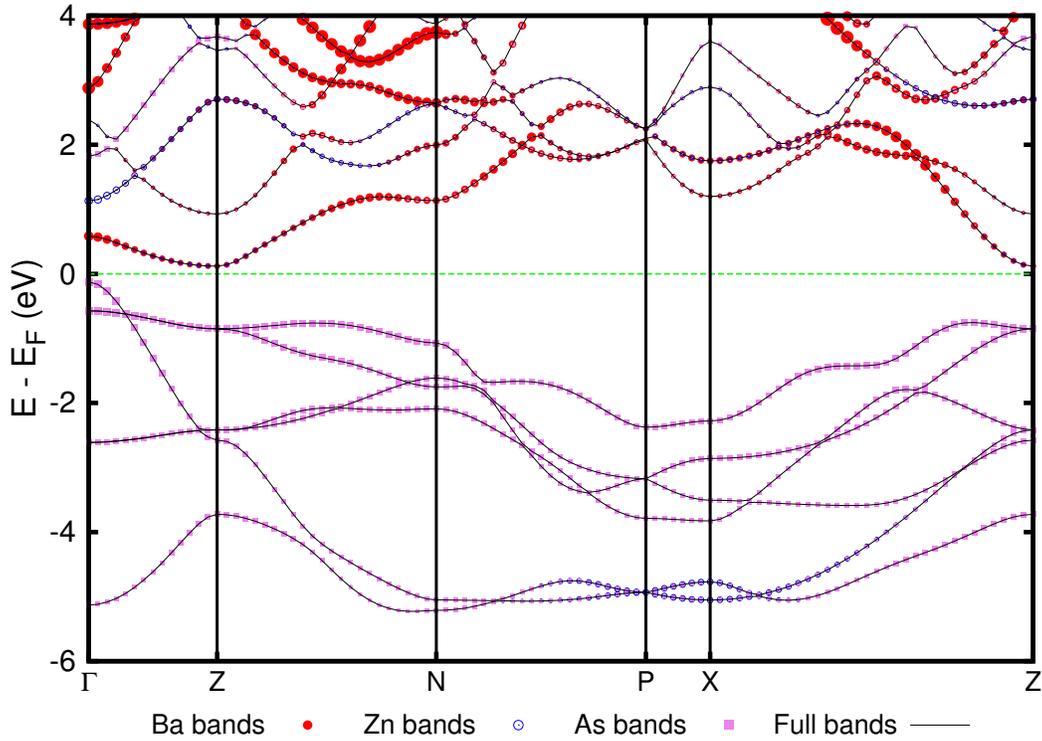}
\end{center}
\caption{(Color online) The character-resolved band structure of BaZn$_2$As$_2$ calculated using
the mBJ functional.} \label{pic-bza-elkbands}
\end{figure*}

\emph{Electronic structure}-The band structure of pure BaZn$_2$As$_2$ calculated
using ELK and the modified Becke-Johnson (mBJ) functional \cite{Becke2006:JCP,tranblaha} is in Fig.~\ref{pic-bza-elkbands}. The band
character is indicated in the legend. The valence band maximum is primarily As
states and the conduction band minimum is primarily Ba states with a small
amount of Zn character. Finally, the LAPW electronic structure for pure
BaZn$_2$As$_2$ is in excellent agreement with that generated by the
pseudopotential method also using the mBJ functional.

\begin{figure*}
\begin{center}
\includegraphics[width=0.60\textwidth]{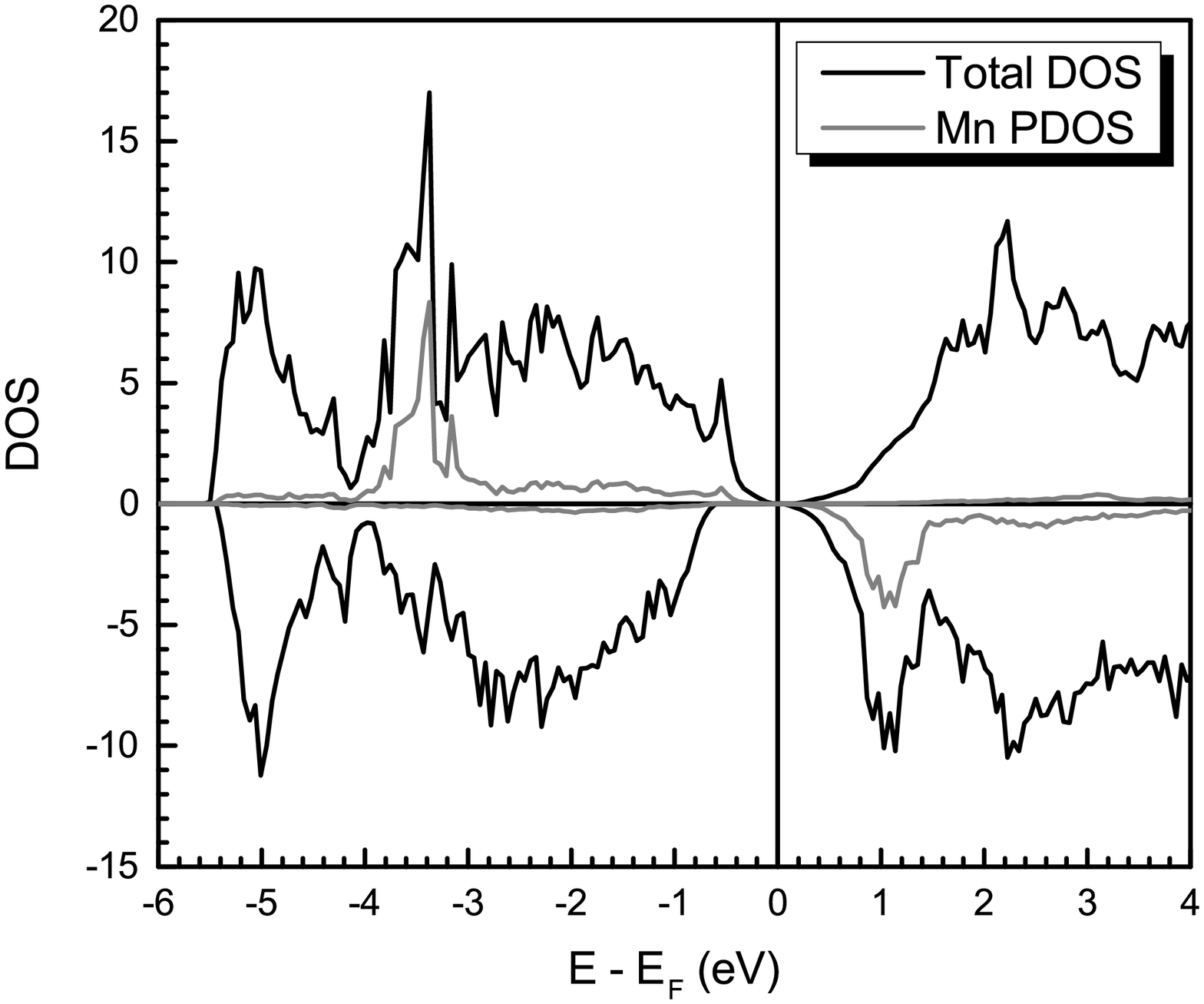}
\end{center}
\caption{The total density of states of Ba(Zn$_{0.875}$Mn$_{0.125}$)$_2$As$_2$
and the partial density of states for Mn calculated using the mBJ functional.}
\label{pic-bza-dos}
\end{figure*}

The density of states (DOS) for BaZn$_2$As$_2$, obtained using VASP and the mBJ functional, is shown
in Fig.~\ref{pic-bza-dos}, where the total DOS and the Mn partial DOS are both
shown.  The partial DOS of Mn confirms that it is in the Mn$^{2+}$ state.

\emph{Details of calculating $J_{x}^{ij}$}-The
Heisenberg Hamiltonian,
\begin{align}
	\label{appendix-eq-heisenberg} H=\sum_{i<j}J_{x}^{ij}{\hat{S}}_{i}\cdot{\hat{S}}_{j},
\end{align}
defines the exchange parameter $J_{x}^{ij}$ for Mn-Mn pairs placed in the Zn
lattice. The Mn atoms are substituted in the two-dimensional Zn lattice as
shown in Fig.~\ref{pic-bza}(b), which for a Mn atom fixed at site 0
indicates which other site is the 1st nearest neighbor, 2nd nearest neighbor,
etc. We constructed supercells of different dimensions and placed two Mn atoms
in the Zn lattice at different sites. Let us define the $1 \times 1 \times 1$
unit cell to be the conventional 10 atom cell with dimensions $a \times a \times
c$ with two Zn layers as depicted in Fig.~\ref{pic-bza}(a). The dimensions of
the supercells we used in our calculations are then summarized in Table
\ref{table-supercell} along with the Mn doping level $y$ from substituting two of
the Zn atoms with Mn atoms.

To extract the exchange parameter, we calculated the energy difference $E_{AFM}
- E_{FM}$ for an AFM and FM alignment of the
Mn moments in VASP for the various supercells. These calculations were carried
out for hole-doping concentrations of $x = 0.0$, $x=0.2$, and $x=0.4$ in the
VCA. The intraplane and interplane neighbors used in our calculations for each
supercell dimension are summarized in Table \ref{table-supercell}, which
refer to the neighbor pairings indicated in Fig.~\ref{pic-bza}(b). The energies
for the intraplane FM and AFM alignments in the $3 \times 3 \times 1$ supercell
are reported in Table \ref{table-energies} for the different hole-doping levels.
Finally, the magnitudes of the extracted exchange parameters and their standard
errors for the different hole-doping levels are summarized in Table
\ref{table-jconstants}. Note that the listed errors do not reflect inaccuracy of the DFT calculations, but rather the standard errors of the fitting procedure. Physically, it corresponds to
the Heisenberg model not being an entirely accurate description of the exchange
interaction in metallic systems.

\begin{table*}
\caption{The supercells dimensions and intraplanar and
interplanar Mn-Mn configurations used for fitting to
Eq.~(\ref{appendix-eq-heisenberg}). Each neighbor number entry represents two calculations, one for
ferromagnetic alignment and another for antiferromagnetic alignment of the Mn
atoms. The doping level $y$ for two Mn atoms placed in the supercells is also
reported.} \label{table-supercell}
\begin{tabular}{|c|c|c|c|}
\hline
Dimensions & $y$ & Intraplane Neighbors & Interplane Neighbors \\ \hline
$2 \times 2 \times 1$ & $0.06250$ & $1, 2, 3, 4, 5$ & $1, 2, 3, 4, 6$ \\ \hline
$3 \times 3 \times 1$ & $0.02778$ & $1, 2, 3, 4, 5, 6, 7$ & \\ \hline
$4 \times 2 \times 1$ & $0.03125$ & $1, 2, 3, 6, 7$ & \\ \hline
$2 \times 2 \times 2$ & $0.03125$ & & $1, 2, 3, 4, 5$ \\ \hline
\end{tabular}
\end{table*}

\begin{table*}
\caption{The distances between Mn-Mn pair configurations and the magnitude and standard error of the exchange constants
for both intraplanar ($J^{\parallel}_{x}$) and interplanar ($J^{\perp}_{x}$) couplings. The pairs are reported by their neighbor
number as shown in the diagram in Fig.~\ref{pic-bza}(b)} \label{table-jconstants}
\begin{tabular}{|c|c|c|c|c|c|c|c|c|}
\hline
 & & \multicolumn{3}{c|}{$J^{\parallel}_{x}(r_{\parallel})$} & & \multicolumn{3}{c|}{$J^{\perp}_{x}(r_{\perp})$} \\
 & & \multicolumn{3}{c|}{(meV)} & & \multicolumn{3}{c|}{(meV)} \\ \hline
\# & $r_{\parallel}$ (\AA) &  $x = 0.0$ & $x = 0.2$ & $x = 0.4$ & $r_{\perp}$ (\AA) & $x = 0.0$ & $x = 0.2$ & $x = 0.4$ \\ \hline
1 & $2.921$ & $200.7 \pm 2.2$ & $108.8 \pm 10.3$ & $60.6 \pm 9.1$ & $6.741$ & $20.9 \pm 2.0$ & $-6.6 \pm 2.7$ & $-20.2 \pm 2.9$\\ \hline
2 & $4.131$ & $81.4 \pm 2.7$ & $-3.1 \pm 12.7$ & $-43.2 \pm 12.7$ & $7.346$ & $4.2 \pm 1.9$ & $-11.8 \pm 2.7$ & $-16.7 \pm 2.9$ \\ \hline
3 & $5.842$ & $-3.2 \pm 1.1$ & $-17.6 \pm 5.1$ & $-13.7 \pm 4.5$ & $7.906$ & $3.2 \pm 0.8$ & $-7.2 \pm 2.7$ & $-9.746 \pm 2.9$ \\ \hline
4 & $6.532$ & $3.1 \pm 1.1$ & $-20.8 \pm 5.4$ & $-25.3 \pm 4.7$ & $8.920$ & $0.9 \pm 0.6$ & $-2.3 \pm 1.1$ & $-3.0 \pm 1.2$ \\ \hline
5 & $8.262$ & $5.3 \pm 0.5$ & $-8.3 \pm 2.4$ & $-15.7 \pm 2.1$ & $9.386$ & $1.2 \pm 1.0$ & $-1.3 \pm 2.8$ & $-4.1 \pm 3.0$ \\ \hline
6 & $8.763$ & $0.4 \pm 1.0$ & $-2.6 \pm 4.8$ & $-6.2 \pm 4.2$ & $10.663$ & $0.3 \pm 0.4$ & $-0.5 \pm 0.6$ & $-1.8 \pm 0.6$ \\ \hline
7 & $9.237$ & $0.6 \pm 0.8$ & $-3.8 \pm 4.0$ & $-2.9 \pm 3.5$ & & & & \\ \hline
\end{tabular}
\end{table*}

\begin{table*}
\caption{The calculated energies of the $3 \times 3 \times 1$ supercell for
intraplane Mn-Mn pair configurations with ferromagnetic (FM) and
antiferromagnetic (AFM) alignment. The pairs are reported by their neighbor
number as shown in the diagram in Fig.~\ref{pic-bza}(b). The three pairs of
columns are the three different hole-doping levels used for $x$.}
\label{table-energies}
\begin{tabular}{|c|c|c|c|c|c|c|}
\hline
 & \multicolumn{2}{c|}{$x = 0.0$} & \multicolumn{2}{c|}{$x = 0.2$} & \multicolumn{2}{c|}{$x = 0.4$} \\
 & \multicolumn{2}{c|}{(eV)} & \multicolumn{2}{c|}{(eV)} & \multicolumn{2}{c|}{(eV)} \\ \hline
\# & FM & AFM & FM & AFM & FM & AFM \\ \hline
1 & -626.8359 & -627.2429 & -428.9581 & -429.1655 & -322.1117 & -322.2600 \\ \hline
2 & -626.9715 & -627.1397 & -429.0902 & -429.0730 & -322.2467 & -322.1584 \\ \hline
3 & -627.0589 & -627.0548 & -429.1219 & -429.0755 & -322.2930 & -322.2575 \\ \hline
4 & -627.0479 & -627.0605 & -429.1160 & -429.0636 & -322.2854 & -322.2248 \\ \hline
5 & -627.0427 & -627.0706 & -429.1178 & -429.0587 & -322.2859 & -322.2181 \\ \hline
6 & -627.0567 & -627.0613 & -429.1129 & -429.1018 & -322.2844 & -322.2843 \\ \hline
7 & -627.0571 & -627.0634 & -429.1134 & -429.0890 & -322.2911 & -322.2754 \\ \hline
\end{tabular}
\end{table*}

\emph{Mn-Mn pair distribution}-In the main text we discuss that the reduction of
the measured moment of Mn in experiment can be understood using a thermodynamic
argument. In short, a statistical calculation of the entropy of the chance of
one Mn atom having another Mn for a neighbor in the thermodynamic limit yields
an expression for the reduction of the total magnetization per Mn atom. Here, we
explicitly derive this expression.

Let there be $m$ Mn atoms that occupy sites on a two-dimensional square lattice
with $n$ sites. The total number of bonds in this system is $2n$ and there are
$k$ Mn-Mn dimers. The total number of possible ways to populate these $2n$ bonds
with $k$ dimers is
\begin{align}
	\label{appendix-eq-binomial1} \binom{2n}{k} &= \frac{(2n)!}{k!(2n-k)!}.
\end{align}
This leaves $m-2k$ Mn atoms to populate the remaining $n-2k$ sites, and the
total number of ways to do this is
\begin{align}
	\label{appendix-eq-binomial2} \binom{n-2k}{m-2k} &= \frac{(n-2k)!}{(n-m)!(m-k)!}.
\end{align}
The total combinations for decorating the square lattice is the product of
Eqs.~(\ref{appendix-eq-binomial1}) and (\ref{appendix-eq-binomial2}), which we
define as $W$. This is correct only in the dilute regime $m \ll n$, as we
neglect instances where any of the remaining $(n-2k)$ Mn atoms ends up next to a
dimer, as well as the possibility that two dimers border each other at a right
angle.

We now approximate the factorials using Stirling's formula,
\begin{align}
	n! &\approx \sqrt{2\pi n}(n/e)^{n},
\end{align}
take the logarithm of $W$ and expand in $1/n$ (again, possible in the dilute limit).
The resulting entropy is
\begin{align}
	S &= \ln \left( \frac{2^{k-1} \left(  \frac{k}{e} \right)^{-k} e^{m-k} e^{k-m} \left( \frac{m-2k}{e} \right)^{2k-m} n^{m-k}}{\pi \sqrt{k} \sqrt{m-2k}}\right).
\end{align}
We now substitute $m=yn,$ $k=\kappa n,$ and write down the free energy $F$ per
site,
\begin{align}
	-\frac{F}{T} &= -\kappa \frac{\Delta E}{T} + \frac{S}{n},
\end{align}
where $\Delta E$ is the energy gained by forming a dimer compared with two isolated Mn
atoms. We now minimize the free energy with respect to the number of dimers: we
expand the logarithm again in $1/n,$ take the derivative with respect to
$\kappa$, and get
\begin{multline}
	\ln \left[ \left( 2 e^{\Delta E/T} \left( -2 \kappa + y \right)^{2} \right) / \kappa \right] \\ = \ln \left[ \left( 2 \beta^{-1} \left( -2 \kappa + y \right)^{2} \right) / \kappa \right] = 0,
\end{multline}
where $\beta=e^{-\Delta E/T}$. Solving for $\kappa$, we get $\kappa = (\beta +8y
-\sqrt{\beta^{2} + 16\beta y})/16$. 

In the main text we established that nearest-neighbor
dimers are AFM, and therefore each dimer results in
the cancellation of two Mn moments from the total magnetization. Therefore the
reduction coefficient (observed magnetization vs. maximum possible) can be
defined as
\begin{align}
	\label{appendix-eq-rcoefficient} r = 1 - 2\beta/y = (\sqrt{\beta^{2} + 16\beta y} - \beta)/8y.
\end{align}

There are a couple of limiting cases that we should note. The first is the case
of infinite energy gain, when $\beta = 0$. As expected, $r=0$ as all Mn atoms
are paired into dimers. The second limiting case is that of no energy gain, when
$\beta = 1$. In this case we should expect the solution to coincide with the
``stochastic'' solution, $r = (1-y)^{4} + 4y^{3} (1-y)$, which is the computed
probability that a given Mn atom forms, stochastically, a dimer with one or a
tetramer with three neighboring Mn atoms (both combinations will cancel the
involved Mn atoms' contribution to the total magnetization). Our solution in
Eq.~(\ref{appendix-eq-rcoefficient}), derived in the limit $y \ll 1$, is correct
to the lowest order in $y$, yielding $r = 1 - 4y +O(y^{2})$. In the next order
our formula underestimates the reduction, $r = 1 - 4y + 32y^{2} + O(y^{3})$ vs.
$r = 1 - 4y + 4y^{2} + O(y^{3})$, and for $y=0.05$ the error is less than $5\%$.

\emph{Stability of interstitial Mn impurities}-A common issue with the III-V
dilute magnetic semiconductor compounds is that Mn doping can lead to both
interstitial and substitutional impurities. Here we show that there is a large
energy penalty for interstitial impurities in BaZn$_2$As$_2$ using the
methodology developed by Jenkins \cite{Jenkins2004:PRB}.

\begin{figure*}
\begin{center}
\includegraphics[width=0.80\textwidth]{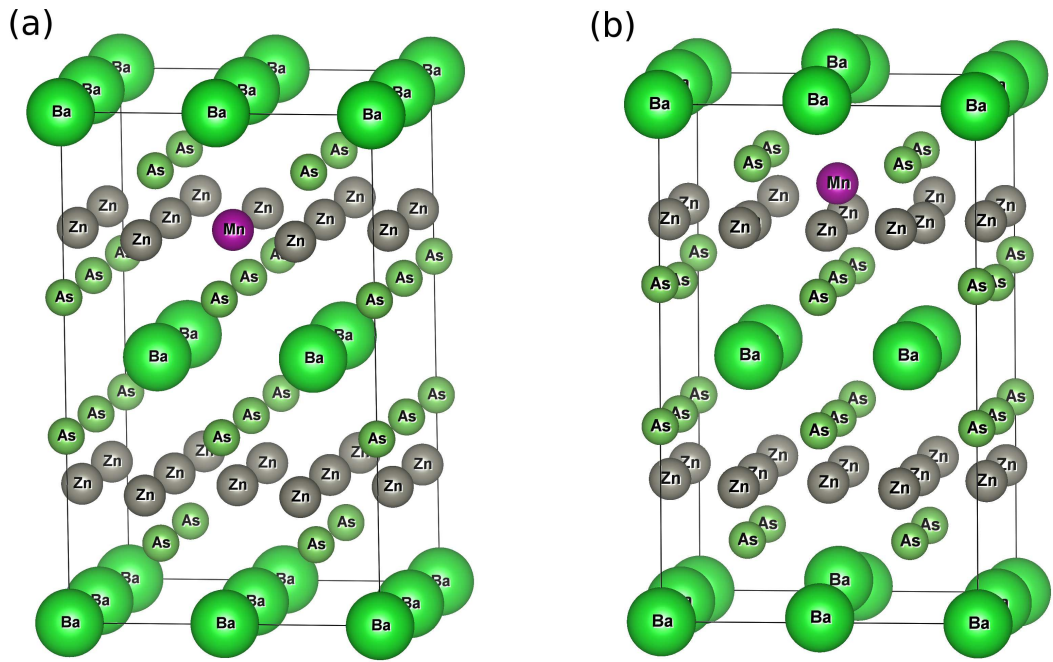}
\end{center}
\caption{The supercell configurations of BaZn$_2$As$_2$ used to calculate the relative energetic stability of a substitutional vs.~interstitial Mn impurity. (a) The relaxed structure of Ba$_{8}$Zn$_{15}$MnAs$_{16}$ where Mn was substituted for a Zn atom. (b) The relaxed structure of Ba$_{8}$Zn$_{16}$MnAs$_{16}$ where the Mn interstitial is in the empty space between neighboring As-Zn planes.} \label{pic-chempot}
\end{figure*}

Let $\mu_{\text{BaZn}_2\text{As}_2}^{\text{bulk}}$ be the bulk chemical potential of BaZn$_2$As$_2$, $\mu_{\text{BaAs}_2}^{\text{bulk}}$ the bulk chemical potential of a
hypothetical structure of BaAs$_2$, and $\mu_{\text{Zn}}^{\text{bulk}}$ the bulk chemical potential of Zn. The potentials of the constituents of BaZn$_2$As$_2$ are subject to the constraint
\begin{align}
	\label{appendix-eq-e0} \mu_{\text{BaZn}_2\text{As}_2}^{\text{bulk}} &= 2 \mu_{\text{Zn}} + \mu_{\text{BaAs}_2}.
\end{align}
The constituents of BaZn$_2$As$_2$ are also related to their respective bulk chemical potentials in the following inequalities
\begin{align}
	\label{appendix-eq-e1} \mu_{\text{BaAs}_2}^{\text{bulk}} &> \mu_{\text{BaAs}_2}, \\
	\label{appendix-eq-e2} \mu_{\text{Zn}}^{\text{bulk}} &> \mu_{\text{Zn}}.
\end{align}
It follows from combining Eqs.~(\ref{appendix-eq-e0}), (\ref{appendix-eq-e1}), and
(\ref{appendix-eq-e2}) that $\mu_{\text{Zn}}$ is subject to the following constraint,
\begin{align}
	\label{appendix-eq-inequality1} \frac{\mu_{\text{BaZn}_2\text{As}_2}^{\text{bulk}} - \mu_{\text{BaAs}_2}^{\text{bulk}}}{2} < \mu_{\text{Zn}} < \mu_{\text{Zn}}^{\text{bulk}}.
\end{align}
In practice the bulk potentials are equivalent to the DFT energy, so to find them we calculate the total energy of bulk BaZn$_2$As$_2$, bulk Zn, and bulk BaAs$_2$ \cite{Jain2013}. 

Next we write down the Gibbs free energy for two Mn impurity scenarios. Let $G_{\text{sub}}$ be
the Gibbs free energy and $E_{\text{sub}}$ be the calculated energy of the supercell with a
substitutional impurity, and let $G_{\text{int}}$ be the Gibbs free energy and $E_{\text{int}}$ be the
calculated energy of the supercell with an interstitial impurity. For our
calculations, the chemical formula for the substitutional impurity supercell is
Ba$_{8}$Zn$_{15}$MnAs$_{16}$ and for the interstitial impurity it is
Ba$_{8}$Zn$_{16}$MnAs$_{16}$, see Fig.~\ref{pic-chempot} for a schematic
representation. The following then is the Gibbs free
energy of the supercells with either a substitutional or interstitial impurity:
\begin{align}
	\label{appendix-eq-e3} G_{\text{sub}} &= E_{\text{sub}} - 15 \mu_{\text{Zn}} - 8 \mu_{\text{BaAs}_2} - \mu_{Mn} \\
	\label{appendix-eq-e4} G_{\text{int}} &= E_{\text{int}} - 16 \mu_{\text{Zn}} - 8 \mu_{\text{BaAs}_2} - \mu_{Mn}
\end{align}
It follows then that the difference in Gibbs free energy between the substitutional
and interstitial configurations is given by
\begin{align}
	\label{appendix-eq-fequation} G_{\text{sub}} - G_{\text{int}} &= E_{\text{sub}} - E_{\text{int}} + \mu_{\text{Zn}}
\end{align}

Using VASP, we
calculate $\mu_{\text{BaZn}_2\text{As}_2}^{\text{bulk}} = -16.98 \text{ eV/f.u.}$, $\mu_{\text{BaAs}_2}^{\text{bulk}} = -13.60 \text{ eV/f.u.}$, $\mu_{\text{Zn}}^{\text{bulk}}
= -1.26 \text{ eV/Zn}$, $E_{\text{sub}} = -143.64 \text{ eV}$, and $E_{\text{int}} = -142.55
\text{ eV}$. We find that the bounding for Zn is $-1.69 \text{ eV} <
\mu_{\text{Zn}} < -1.26 \text{ eV}$ and that $G_{\text{sub}} - G_{\text{int}} = -1.09 \text{ eV} +
\mu_{\text{Zn}}$. Based on this analysis, there is a substantial energetic
preference of $-2.6 \pm 0.2 \text{eV}$ for the substitutional impurity.

\bibliography{bza}

\begin{thebibliography}{40}
\bibitem [{par()}]{partition}%
  \BibitemOpen
  \href@noop {} {}\bibinfo {note} {The difference is due to different space
  partitioning in different codes.}\BibitemShut {Stop}%
\bibitem [{hop()}]{hop}%
  \BibitemOpen
  \href@noop {} {}\bibinfo {note} {While the relevant hopping amplitudes
  $t_{pd}$ differ from orbital to orbital, for our purpose introducing one
  effective parameter is sufficient.}\BibitemShut {Stop}%
\bibitem [{map()}]{map}%
  \BibitemOpen
  \href@noop {} {}\bibinfo {note} {Magnetic interactions in itinerant systems
  do not always follow the Heisenberg law, so the mapping is never exact, but
  it gives us a good visual picture.}\BibitemShut {Stop}%
\end{thebibliography}

\end{document}